\documentclass[prd,nofootinbib]{revtex4}
\pdfoutput=1
\usepackage{amsmath,amssymb}
\usepackage{cancel,graphicx}
\begin{document}

\title{Minimal Flavor Protection:\\ A New Flavor Paradigm in Warped
  Models} 

\author{Jos\'e Santiago}
\email{santiago@itp.phys.ethz.ch}
\affiliation{Institute for Theoretical Physics, ETH, CH-8093,
  Z\"urich, Switzerland}

\begin{abstract}
We propose a new flavor paradigm for models with warped extra
dimensions. The idea is to impose the
minimal amount of flavor protection to make warped models compatible
with all current flavor and electroweak precision constraints.
We discuss a particular realization of this minimal flavor protection 
in the quark sector, 
by means of a flavor symmetry acting on the right handed down sector. 
Hierarchical quark masses and mixing angles are
naturally reproduced through wave function localization, 
and flavor violating processes are predicted, in the absence of large brane
kinetic terms for the right handed down quarks,
below but not too far from current experimental limits 
in several channels.  With this new flavor pattern, models with warped
extra dimensions can be accessible through direct production of new
resonances at the LHC and also through precision flavor experiments. 
\end{abstract}

\maketitle

\section{Introduction} 

Models with warped extra dimensions~\cite{Randall:1999ee} provide a neat
explanation to the vast hierarchy between the Planck and electroweak
scales, with the added bonus of a very appealing structure of
flavor. First, the exponential localization of the chiral zero
modes of bulk fermions~\cite{Grossman:1999ra} 
naturally predicts a hierarchical structure of
quark masses and mixing angles~\cite{Gherghetta:2000qt,Huber:2000ie}.
Even more important, the effect of the warp
factor on the localization of the gauge boson Kaluza-Klein (KK) modes
provides a very special flavor protection, dubbed the
RS-GIM mechanism~\cite{Gherghetta:2000qt,Huber:2003tu,Agashe:2004cp}.
Thanks to this mechanism, flavor violating processes are suppressed by
small mixing angles or quark masses. 
Thus, flavor violating processes involving light fermions,
which are very constrained experimentally, are naturally small.
In most of the studies of models with warped extra dimensions, it has
been implicitly assumed that, either flavor physics is safe,
thanks to the RS-GIM mechanism, or that possible flavor violations
above experimental limits can be fixed (with some fine-tuning or with
appropriate flavor symmetries) without altering the natural
generation of hierarchical quark masses and mixing angles and without
affecting flavor diagonal physics in a sensitive way. With that
assumption, realistic models with a low scale of new physics have been
recently constructed~\cite{Carena:2006bn}, using ideas like custodial 
symmetry~\cite{Agashe:2003zs} and a protection of the $Zb_L\bar{b}_L$
coupling~\cite{Agashe:2006at,Djouadi:2006rk}. 
The result is that KK excitations of
gauge bosons as light as 
$M_{KK} \gtrsim 3~\mathrm{TeV}$ 
can be compatible with electroweak precision tests (EWPT) and a
natural explanation of the fermion mass hierarchy.

Detailed analyses of flavor
constraints~\cite{Agashe:2004cp,Huber:2003tu,Moreau:2006np,newflavor,neubert} 
show that the RS-GIM mechanism is in fact extremely effective (see
also~\cite{Davidson:2007si}), predicting
flavor violating processes close to, but below current experimental bounds
in almost all channels. The only notable exception is CP violation in
the Kaon 
system, $\epsilon_K$, which, if \textit{simultaneous
flavor violating left handed (LH) and right handed (RH) 
currents} mediated by the gluon KK modes are present, requires
the mass of the lightest KK mode to be (depending on the particular
model)~\cite{newflavor} 
\begin{equation}
M_{KK} \gtrsim 20-30~\mathrm{TeV}, \quad \mbox{(from $\epsilon_K$, LR
  contribution)}. 
\end{equation}

The main goal of this note is to emphasize that, despite recent
claims that models with warped extra dimensions suffer from a serious
flavor problem, the above discussion shows that the situation is
actually much better than that.
As we just said, 
the RS-GIM mechanism is strikingly effective, seriously missing only
in one observable and due to a very particular chirality enhanced
contribution. Our proposal is to make use of the already good flavor
properties of models with warped extra dimensions and introduce only the 
minimal amount of flavor protection to suppress the dangerous
contribution to $\epsilon_K$, without sizably modifying the remaining 
flavor properties of the model.~\footnote{Less minimal 
  proposals of flavor protection in models
  with warped extra dimensions
  have been presented
  in~\cite{Cacciapaglia:2007fw}. See~\cite{Davoudiasl:2008hx} for a
model that is compatible with flavor constraints at the expense of not
fully explaining the hierarchy problem.} 
This \textbf{minimal flavor protection (MFP)} prescription 
can be realized, for instance, by imposing a $U(3)$ flavor symmetry
under which all fields are singlets except for
the three fields that give rise to the charge $-1/3$ RH
quark zero modes (and other fields related to them by the
symmetries of the model), which transform as a triplet. The symmetry is only
broken by brane localized Yukawa couplings (or by localized mass
or kinetic mixing for the multiplets of the brane gauge symmetry 
that do not contain the $d_R$ zero modes) 
generating in this way non-trivial quark masses in the down
sector.
As we will see in detail below,
this new paradigm still allows for the quark masses and mixing angles
to be induced by wave function localization. The masses in
the up sector are generated by double hierarchies, whereas the ones in
the down sector are generated by a single hierarchy and therefore are
naturally less spread. This provides a rationale for the smaller
hierarchy of quark masses in the down sector ($m_d/m_b\sim
10^{-3}$) than in the up sector ($m_u/m_t \sim 10^{-5}$).
Also, LH flavor violating currents mediated by the gluon KK modes
are predicted as
in the standard realization of flavor (below but not too far from
current experimental bounds) and therefore could be measured in the
near future, particularly observables in the $B$ system~\cite{Agashe:2004ay}.  
The plan of the paper is the following: In the next section we
describe in detail a particular realization of the MFP proposal in the
quark sector, discussing the absence of the leading contribution to $\Delta
F=2$ left-right (LR) processes from dimension 6 operators. 
In section~\ref{other:sources} we discuss sub-leading (higher dimension
or loop suppressed operators) flavor violating contributions which
give the main
constraint in MFP models. We discuss the numerical significance of the
corresponding bounds in section~\ref{section:scan} and finally
conclude with a discussion of the results and an outlook of future
prospects in section~\ref{discussion}.

\section{Minimal Flavor Protection in the Quark Sector\label{MFP}}

In this section we describe in detail an explicit realization of
MFP in the quark sector. For the sake of the presentation we
consider a model with a 
fundamental Higgs localized at the IR brane and discuss the
results for other models in section~\ref{section:scan}. 
The discussion closely follows the one in~\cite{newflavor}.
The background is a slice of AdS$_5$,
\begin{equation}
ds^2=\left(\frac{R}{z}\right)^2 ( \eta_{\mu\nu} dx^\mu dx^\nu - dz^2),
\end{equation}
with $R \leq z \leq R^\prime$. $R\approx M_\mathrm{Pl}^{-1}$ and
$R^\prime \approx \mathrm{TeV}^{-1}$ are the position of the UV and IR
branes, respectively. 
In order to make it compatible with a low KK scale we assume a bulk
$SU(3)\times SU(2)_L \times SU(2)_R \times U(1)_X$ gauge
group~\cite{Agashe:2003zs} and a discrete $L \leftrightarrow R$
symmetry that exchanges the $SU(2)_L$ and $SU(2)_R$ groups, to protect
the $Z\bar{b}_L b_L$ coupling~\cite{Agashe:2006at}. 
We introduce three families of five-dimensional bulk quarks
with the following quantum numbers under $SU(2)_L \times
SU(2)_R \times U(1)_X$,
\begin{equation}
Q^i=(2,2)_{2/3}, \quad Q^{ui}=(1,1)_{2/3},
\quad Q^{di}=(1,3)_{2/3},\quad \tilde{Q}^{di}=(3,1)_{2/3},
\end{equation} 
where $i=1,2,3$ is the flavor index. The choice of boundary conditions
is 
\begin{equation}
Q^i = \begin{pmatrix} 
\chi^i=\begin{pmatrix}\chi^{ui} \\
  \chi^{di}\end{pmatrix} [-+] 
\\ 
q^i=\begin{pmatrix}q^{ui} \\
  q^{di}\end{pmatrix} [++] 
\end{pmatrix},
\quad
Q^{ui}=U^{i} [--],
\quad
Q^{di} = \begin{pmatrix} X^i[+-] \\ U^{\prime\,i}[+-] \\ D^i[--]
\end{pmatrix}, 
\quad
\tilde{Q}^{di} = \begin{pmatrix} \tilde{X}^i[+-] \\
  \tilde{U}^{\prime\,i}[+-] \\ \tilde{D}^i[+-], 
\end{pmatrix}, 
\end{equation}
where $\chi^i$ and $q^i$ are $SU(2)_L$ doublets with hypercharge, $7/6$
and $1/6$, $\tilde{Q}^{di}$ forms a triplet under $SU(2)_L$ and 
$X^i,U^{\prime\,i}$ and $D^i$ are the three
components of the $SU(2)_R$ triplet, with $T^3_R=+1,0,-1$,
respectively. 
$-$ ($+$) denotes Dirichlet boundary condition for the LH (RH)
chirality of the bulk fermion at the corresponding brane (the first
sign is for the UV brane and the second for the IR brane).
A LH (RH) fermion zero mode corresponds to a field with $[++]$
$([--])$ boundary conditions. With our choice of boundary conditions,
$q^i_L$, $U^i_R$ and $D^i_R$ are the only fields that have zero modes. 
The multiplet $\tilde{Q}^{di}$ has been included to
ensure that the heavy physics is approximately invariant under the
discrete $L \leftrightarrow R$ symmetry~\cite{Agashe:2006at}.  
In order to realize our MFP prescription, we assume a global $U(3)$
symmetry under 
which $Q^{di}$ and $\tilde{Q}^{di}$ transform as triplets and all the
other fields transform as singlets. 

Bulk fermions admit a bulk mass term that can always be
taken to be diagonal in flavor space. Using standard notation we
parametrize these mass terms as
\begin{equation}
\left(\frac{R}{z}\right)^4 \Big[ 
\frac{c_{qi}}{z} \bar{Q}^{i} Q^{i}
+\frac{c_{ui}}{z} \bar{Q}^{ui} Q^{ui}
+\frac{c_d}{z} 
\big(
\bar{Q}^{di} Q^{di}
+\bar{\tilde{Q}}^{di} \tilde{Q}^{di}
\big)
\Big],
\end{equation}
where we have explicitly written a common bulk mass for all $Q^{di}$
and $\tilde{Q}^{id}$ as
imposed by the $U(3)$ flavor symmetry and the discrete $L
\leftrightarrow R$ symmetry.
These bulk fields admit zero modes given by
\begin{eqnarray}
q_L^{i}(x,z)&=&\chi_{c_{qi}}(z) q^{(0)i}_L(x) + \ldots,
\\
U_R^{i}(x,z)&=&\chi_{-c_{ui}}(z) u^{(0)i}_R(x) + \ldots,
\\
D_R^{i}(x,z)&=&\chi_{-c_{d}}(z) d^{(0)i}_R(x) + \ldots,
\end{eqnarray}
with wave function that results in canonical normalization of the
four-dimensional fields given by
\begin{equation}
\chi_c(z) \equiv 
\frac{1}{\sqrt{R^\prime}} 
\left(\frac{z}{R}\right)^2 \left(\frac{z}{R^\prime}\right)^{-c}
f_{c},
\end{equation}
with
\begin{equation}
f_c\equiv\sqrt{\frac{1-2c}{1-\left(\frac{R^\prime}{R}\right)^{2c-1}}}.
\label{fc} 
\end{equation}
The fermion zero
mode masses come from $U(3)$ violating IR localized Yukawa couplings.
After EWSB, they read,
\begin{eqnarray}
\mathcal{L}_y&=&
-\frac{v}{\sqrt{2}} \left(\frac{R}{R^\prime}\right)^4
R^\prime \big[(\bar{q}^{ui}-\bar{\chi}^{di})\tilde{Y}^u_{ij} U^j 
+ \bar{q}^{di} \tilde{Y}^d_{ij} (D^j+\tilde{D}^j) +
\mathrm{h.c.}+\ldots\big]
\\
&=&
-\frac{v}{\sqrt{2}} (\bar{u}_L^{(0)i} f_{qi}\tilde{Y}^u_{ij}
f_{-uj}u_R^{(0)j}  
+ \bar{d}_L^{(0)i} f_{qi} \tilde{Y}^d_{ij} f_{-d}d_R^{(0)j}) +
\mathrm{h.c.} + \ldots~. \label{yukawa:lag} \nonumber
\end{eqnarray} 
We have written the dimensionful five-dimensional Yukawa couplings as
$Y_{5D}^{u,d}= R^\prime \tilde{Y}^{u,d}$, where the (now
dimensionless) Yukawa couplings $\tilde{Y}^{u,d}$ are assumed to be
anarchic $3 \times 3$ matrices 
(all entries order one and order one determinant). In the
second line we have explicitly written the effective four-dimensional
Yukawa couplings for the quark zero modes. $v=246$ GeV is the warped
down Higgs vev.
The hierarchical structure of the quark masses and mixing angles is
then explained by a hierarchical structure of the $f_{qi}$ and
$f_{-ui}$~\cite{Huber:2003tu} (recall that the $U(3)$ flavor symmetry
forces a common $f_{-d}$),   
which are generated by non-hierarchical five-dimensional bulk masses,
see Eq.~(\ref{fc}), 
\begin{equation}
f_{q1} \ll f_{q2} \ll f_{q3}, 
\quad
f_{-u1} \ll f_{-u2} \ll f_{-u3}. 
\end{equation}
The SM quark masses can then be diagonalized with unitary rotations,
\begin{eqnarray}
&&\frac{v}{\sqrt{2}} (\mathcal{U}_L^\dagger f_q \tilde{Y}_u f_{-u}
  \mathcal{U}_R)_{ij} =
m^u_i \delta_{ij}, \\ 
&&\frac{v}{\sqrt{2}} (\mathcal{D}_L^\dagger f_q \tilde{Y}_d f_{-d}
  \mathcal{D}_R)_{ij} =
m^d_i \delta_{ij},  \label{massdiagonalization}
\end{eqnarray}
which, due to the hierarchical structure of $f_{q,-u}$, are
hierarchical in the case of $\mathcal{U}_{L,R}$ and $\mathcal{D}_L$,
\begin{equation}
|(\mathcal{U}_L)_{ij}| \sim |(\mathcal{D}_L)_{ij}| \sim \frac{f_{qi}}{f_{qj}}, 
\quad
|(\mathcal{U}_R)_{ij}| \sim \frac{f_{-ui}}{f_{-uj}},\label{rotationmatrices}
\quad i \leq j,
\end{equation}
and therefore the CKM matrix is also hierarchical,
\begin{equation}
|(V_{CKM})_{ij}|=|(\mathcal{U}_L^\dagger \mathcal{D}_L)_{ij}| 
\sim \frac{f_{qi}}{f_{qj}}, \quad i \leq j.
\end{equation}
From here on, the $\sim$ symbol means that the equalities are true up
to Yukawa dependent order one numbers. 
On the other hand, 
$\mathcal{D}_R$ is 
the order one unitary matrix that diagonalizes the
matrix 
\begin{equation}
(\tilde{Y}^d)^\dagger_{ik}f_{qk}^2\tilde{Y}^d_{kj}.
\end{equation}
The diagonal masses are then also hierarchical,
\begin{equation}
m^u_i \sim \frac{v}{\sqrt{2}} f_{qi} f_{-ui},
\quad
m^d_i \sim \frac{v}{\sqrt{2}} f_{qi} f_{-d}.\label{diagonal:masses}
\end{equation}
Assuming that the hierarchical pattern of quark masses and mixing
angles comes from wave function localization and not due to
hierarchies in the fundamental Yukawa couplings (as we have just
discussed above), the CKM matrix and
the up type quark masses approximately fix the values
of the following localization parameters,
\begin{eqnarray}
f_{q3}&\sim& 1, \quad 
f_{q2}\sim \lambda^2,
\quad
f_{q1}\sim \lambda^3, \label{fqi}\\
f_{-u3}&\sim& 1, \quad 
f_{-u2}\sim \frac{m_c}{m_t}\lambda^{-2}, \quad 
f_{-u1}\sim \frac{m_u}{m_t}\lambda^{-3}, \quad \label{fui} 
\end{eqnarray}
where $\lambda\sim 0.22$ is the Cabbibo angle. The localization
parameter in the RH down sector is common to all three flavors, giving
a ratio of masses
\begin{equation}
m_d/m_s/m_b\approx [(0.03-0.1)/1/(50-70)]^\mathrm{exp} 
\sim [0.22/1/21]^\mathrm{MFP}, \label{md:ratios}
\end{equation}
where the first set of numbers is the experimental ratios of masses
(with the variation indicating the uncertainty) at the scale of new
physics $\sim 3$
TeV and the second set is the ratio of masses we obtain (as usual up to
order one Yukawa couplings) from Eq.~(\ref{diagonal:masses}) and the
values of $f_{qi}$ from Eq.~(\ref{fqi}). 
Note that, thanks to the smaller mass hierarchy in the down sector,
the numerical differences can be easily accounted for by a mild hierarchy
in the order one Yukawa couplings. This could not have been achieved
with universality of either $f_q$ or $f_{-u}$, since the hierarchy in
the up sector is too large to be accounted for without hierarchical
Yukawa couplings in that case.

We can now turn to the flavor violation in this model.
In the current eigenstate basis, in which Yukawa couplings are
non-diagonal, the coupling of the quark zero modes to the gauge boson
KK modes, after integration over the extra dimension, is diagonal but
flavor dependent, except for $d_R^{(0)i}$, for which the $U(3)$
symmetry guarantees flavor independence,
\begin{equation}
\bar{u}_L^{(0)i} g^{(n)}_{qi} \cancel{G}^{(n)} u_L^{(0)i}
+\bar{u}_R^{(0)i} g^{(n)}_{ui} \cancel{G}^{(n)} u_R^{(0)i}
+\bar{d}_L^{(0)i} g^{(n)}_{qi} \cancel{G}^{(n)} d_L^{(0)i}
+\bar{d}_R^{(0)i} g^{(n)}_{d} \cancel{G}^{(n)} d_R^{(0)i},
\end{equation}
where we have written the coupling to KK gluons, $G^{(n)}_\mu \equiv
T^a G^{(n)a}_\mu$, which is the largest
one. The
physical basis, with diagonal quark masses, is obtained by the unitary
rotations in
Eq.~(\ref{massdiagonalization}). The flavor dependence of the
couplings to the KK gluons induce flavor changing neutral currents
(FCNC) upon such rotations~\cite{Delgado:1999sv}, with couplings
\begin{eqnarray}
(g^{(n)}_{uL})_{ij}&=& g^{(n)}_{qk} (\mathcal{U}_L)^\ast_{ki}
(\mathcal{U}_L)_{kj}, \quad
(g^{(n)}_{dL})_{ij}= g^{(n)}_{qk} (\mathcal{D}_L)^\ast_{ki}
(\mathcal{D}_L)_{kj}, \nonumber \\
(g^{(n)}_{uR})_{ij}&=& g^{(n)}_{uk} (\mathcal{U}_R)^\ast_{ki}
(\mathcal{U}_R)_{kj},\quad
(g^{(n)}_{dR})_{ij}= g^{(n)}_{d} \delta_{ij}. \label{gluonFCNC}
\end{eqnarray}
Our MFP prescription 
guarantees flavor independence of the couplings of the RH
down quarks to the gauge boson KK modes, and therefore no flavor
violating RH currents are generated in the down sector (due to gauge
KK mode exchange). 

Although the effect of the full gluon KK tower can be easily taken into
account~\cite{newflavor,neubert}, it is instructive to look at the
couplings to the first KK mode to understand the RS-GIM mechanism in
action. The coupling to
the first gluon KK mode is given, to a very good approximation,
by~\cite{newflavor} 
\begin{equation}
g_x \approx g_{s\ast}\left( - \frac{1}{\log R^\prime/R} + f_x^2
\gamma(c_x)\right),
\end{equation}
where here $x$ denotes the corresponding $qi$, $-ui$ or $-d$,
$g_{s\ast}\approx 6$ is the bulk QCD gauge coupling (the actual
four-dimensional coupling is, in the absence of large brane kinetic
terms, $g_{s\ast}/\sqrt{\log (R^\prime/R)}$)
and $\gamma(c)$ is
an order one function. The first term is flavor
universal and does not induce any FCNC. The second,
however is flavor dependent and generates FCNC as shown in
Eq.~(\ref{gluonFCNC}). Inserting the rotation matrices,
Eq.~(\ref{rotationmatrices}), we obtain
\begin{equation}
(g_{uL,dL}^{(1)})_{ij} \sim g_{s\ast} f_{qi} f_{qj},
\quad
(g_{uR}^{(1)})_{ij} \sim g_{s\ast} f_{-ui} f_{-uj},
\quad
(g_{dR}^{(1)})_{ij} = g^{(1)}_d \delta_{ij}.
\end{equation}
This shows the RS-GIM and the MFP mechanisms at work. 
The left handed FCNC are suppressed by ratios of the CKM entries, 
see Eq. (\ref{fqi}),
whereas the up type right handed FCNC are suppressed by ratios of the light
to heavy quark masses, Eq. (\ref{fui}). Finally, the RH down type FCNC are
absent due to the flavor symmetry. Although we have exemplified this
effect with the coupling to the first gluon KK mode, the protection
extends to all the modes and to all gauge bosons. 
In the above discussion, we have neglected brane localized operators.
Brane localized mass terms, for instance, can violate the MFP
prescription, as they can mix the $d_R^{(0)i}$
 with the heavy
$q^{(n)i}$, which have family dependent couplings to the gauge boson
KK modes. Thus, if the fermion quantum numbers of the fields allow for
such mixing, the flavor symmetry
should remain unbroken, on the corresponding brane, for the
multiplets of the surviving symmetry in that brane that contain the
$d_R^{(0)i}$. Brane kinetic terms~\cite{BKT} for $Q^{di}$, on the
other hand, cannot be forbidden by the flavor symmetry, as it is
explicitly broken at the brane for these fields. In particular, they
induce non-universal couplings of the $d_R^{(0)i}$ to the KK gluons at
the IR brane, which in turn, generate FCNC in the right handed down
sector in the physical basis. This effect can be however
parametrically suppressed by a loop factor and will be discussed in
detail in section~\ref{other:sources}.


The flavor violating couplings to the gauge boson KK modes induce
flavor violating four-fermion interactions, once the heavy fields are
integrated out. We take as an example the $\Delta S=2$ contribution to
$\Delta m_K$ and $\epsilon_K$, although all other $\Delta F=2$
processes are similar. 
The relevant operator has the following structure,
\begin{equation}
\frac{1}{2M_{KK}^2} \Big[
\big(
(g_{dL}^{(1)})_{ds}
\bar{d}_L T^a \gamma^\mu s_L
+(g_{dR}^{(1)})_{ds}\bar{d}_R T^a \gamma^\mu s_R
\big)
\big(
(g_{dL}^{(1)})_{ds}\bar{d}_L T^a \gamma_\mu s_L
+(g_{dR}^{(1)})_{ds}\bar{d}_R T^a\gamma_\mu s_R
\big)
\Big],
\end{equation}
where $T^a$ are the color matrices in the fundamental representation, 
$d_{L,R}$ and $s_{L,R}$ denote the physical down and
strange quarks and we have included the first gluon KK mode 
(with mass denoted by $M_{KK}$) for illustration. These dimension 
6 operators can be put in the standard
basis by using Fierz identities and properties of the color matrices
to get a Hamiltonian,
\begin{equation}
\mathcal{H}^{\Delta S=2}= 
C_1^{sd} Q_1^{sd} +\tilde{C}_1^{sd} \tilde{Q}_1^{sd}
+C_4^{sd}Q_4^{sd}+C_5^{sd}Q_5^{sd},
\end{equation}
where the coefficients read,
\begin{eqnarray}
C_1^{sd}&=&\frac{1}{6} \frac{1}{M_{KK}^2}
(g_{dL}^{(1)})_{ds}^2,
\quad \quad \quad \quad
\tilde{C}_1^{sd}=\frac{1}{6} \frac{1}{M_{KK}^2}
(g_{dR}^{(1)})_{ds}^2,
\\
C_4^{sd}&=&-\frac{1}{M_{KK}^2}
(g_{dL}^{(1)})_{ds}
(g_{dR}^{(1)})_{ds}, 
\quad 
C_5^{sd}=\frac{1}{3}\frac{1}{M_{KK}^2}
(g_{dL}^{(1)})_{ds} (g_{dR}^{(1)})_{ds},
\end{eqnarray}
and we have used standard notation for the operators ($\alpha$ and
$\beta$ are color indices)
\begin{eqnarray}
Q_1^{sd}&=&\bar{d}^\alpha_L \gamma^\mu s^\alpha_L 
\bar{d}^\beta_L \gamma^\mu s^\beta_L, 
\quad  
\tilde{Q}_1^{sd}=\bar{d}^\alpha_R \gamma^\mu s^\alpha_R 
\bar{d}^\beta_R \gamma^\mu s^\beta_R, 
\\
Q_4^{sd}&=& \bar{d}^\alpha_R s^\alpha_L \bar{d}^\beta_L s^\beta_R,
\quad \quad \quad
Q_5^{sd}= \bar{d}^\alpha_R s^\beta_L \bar{d}^\beta_L s^\alpha_R.
\end{eqnarray}
Our MFP prescription
ensures that the coefficients of the operators that involve RH currents,
$\tilde{Q}_1^{sd}$, $Q_4^{sd}$ and $Q_5^{sd}$, are zero at this order.

Using the results of~\cite{Bona:2007vi}, it was shown in Ref.~\cite{newflavor} 
that the RS-GIM mechanism is enough to suppress almost
all flavor violating observables below current experimental limits
(but not too far below, in particular in observables in the $B$
system). The only observable that gives a
constraint on the KK excitation of the gauge bosons
stronger than the one obtained from EWPT is the
measurement of CP violation in the Kaon system, $\epsilon_K$. 
Furthermore, the constraint is significant only 
if there are \textit{simultaneous flavor
  violations in 
  both left and right currents}. The reason is that the effect of $\Delta S=2$
operators with both chiralities has an enhancement as compared with
the ones that only involve one chirality proportional to
\begin{equation}
\frac{3}{4} \left(\frac{m_K}{m_s(\mu_L)+m_d(\mu_L)}\right)^2
\eta_1^{-5} \approx 140,
\end{equation}
where $\eta_1$ comes from the RGE running.
Therefore only the (imaginary parts of) the coefficients
$C_{4,5}^{sd}$ are strongly constrained.
Our MFP prescription guarantees, 
in the absence of large BKT, the vanishing of FCNC in RH
down currents
and therefore that $\tilde{C}_1^{sd}=C_{4,5}^{sd}=0$.
Once the LR contributions are absent,
all the other observables are typically less constraining than EWPT.

\section{Extra Sources of Flavor Violation\label{other:sources}}

In the previous section we have seen how, in the absence of BKTs, 
our MFP prescription prevents
the appearance of FCNC RH currents in the down sector, from dimension 6
operators generated by the exchange of gauge boson KK modes. 
Then, the only flavor observable which induces a strong experimental 
constraint does not
receive the LR chirality enhanced contribution that makes it
dangerous. Let us now discuss other sources of FCNC that are present
in our MFP model. 
Although they are all formally sub-leading, as they correspond to
dimension 8 operators or are loop suppressed, 
the absence of the leading contribution makes them the main source of
flavor violation and therefore we have to consider their effect.
We discuss them in turn.
\subsection{Mass Mixing with Kaluza-Klein Modes} 
These new effects can have two different origins. The first is
the mixing of the fermion zero modes with their (vector-like) KK
excitations, which induces in general FCNC for the $Z$ boson and for any
gauge boson KK mode (including the KK gluons). The second is
the mixing of the $Z$ with its KK excitations, if they had FCNC
couplings to the fermion zero modes. 
This latter effect does not occur for the RH down
quark zero modes as, due to flavor universality, they do not have FCNC
with the $Z$ KK modes. Furthermore, 
the corrections to the LH currents, although \textit{a priori} 
flavor violating, have a very effective suppression due
to the mechanism in~\cite{Agashe:2006at} and we have checked that they
are indeed negligibly small.
The effect of the vector-like excitations of the quarks on the
couplings to the $Z$ boson have been computed for models with warped
extra dimensions in~\cite{delAguila:2000kb}, 
using the general results of~\cite{delAguila:2000rc}. The outcome is
that flavor violating couplings of the $Z$ to the RH down
quarks are suppressed by the RS-GIM mechanism while couplings to the LH
down quarks are \textit{forbidden} by the MFP mechanism. The reason is
that the dimension 6 contribution from the mixing with vector-like
quarks comes from the effects of doublets ($q^{(n)}$) for RH currents
and singlets ($d^{(n)}$) for LH currents and therefore the latter are
the ones that the MFP mechanism protects.~\footnote{The correction to
  the $Z \bar{d}_L s_L$ enters at dimension 6 but $\Delta S=2$
  processes require two of these vertices and are therefore dimension 8.} 
Thus, the relevant constraints from dimension 8 operators are expected
to come from the exchange of KK gluons, which we consider now.

The Yukawa mixing of $d_R^{(0)i}$ with the KK excitations of $q^i$
induce flavor violations in the coupling of $d_R^{(0)i}$ to the gluon
KK modes. Using the mass insertion approximation (in the numerical
scans that we discuss below we have included exactly the effect of
fermion mixing by numerically diagonalizing the corresponding mass
matrices), we obtain, for the
coupling of $d^{(0)i}_R$ to the first KK gluon
\begin{equation}
(g^{(1)}_{dR})_{ij}\approx g_{s\ast} 
f_{-d}^2 \bigg[
\mathcal{D}_R^\dagger (\tilde{Y}^d)^\dagger
f_{q}^{(1)}\left(\frac{v }{\sqrt{2}M_{q(1)}}\right)^2
\tilde{Y}^d \mathcal{D}_R \bigg]_{ij}\sim g_{s\ast} f^2_{-d}
\left(\frac{v}{\sqrt{2}M_{KK}}\right)^2,\label{mass:insertion}
\end{equation}
where we have included for illustration only the first fermion KK
mode, $q^{(1)i}$, with a mass denoted in matrix form 
by $M_{q(1)}$ and we have denoted with $f_q^{(1)}$ the 
diagonal matrix containing the (order one) 
couplings of the $q^{(1)i}$ 
to the first gluon KK mode. In the last equality we
have assumed that the fermion KK modes have masses similar to the ones
of the gluon KK modes. (In our numerical scans we have not made use of
that assumption but have included the
exact mass of the fermion KK modes.) 
Thus, after EWSB, we do have a LR contribution to $\epsilon_K$ with
coefficient,
\begin{equation}
C_{4}^{sd(\mathrm{MFP})}\sim
\frac{g_{s\ast}^2}{M_{KK}^2}\left(\frac{v}{\sqrt{2}M_{KK}}\right)^2 
f_{q1}f_{q2} f_{-d}^2, 
\end{equation}
to be compared with the one that appears in the standard realization
of flavor in models with warped extra dimensions
\begin{equation}
C_{4}^{sd(\mathrm{RS})}\sim \frac{g_{s\ast}^2}{M_{KK}^2} f_{q1}f_{q2}
f_{-d1}f_{-d2}.  
\end{equation}
Thus, there is a relative suppression with respect to the standard
result of order
\begin{equation}
\frac{C_4^{sd(\mathrm{MFP})}}{C_4^{sd(\mathrm{RS})}} \sim 
\left(\frac{v}{\sqrt{2}M_{KK}}\right)^2 \frac{f_{-d}}{f_{-d1}}
= \left\{
\begin{array}{l} 
2.5\times 10^{-2} \left(\frac{3~\mathrm{TeV}}{M_{KK}}\right)^2
\frac{f_{-d}/f_{-d1}}{0.22/0.03}, \\
7.\times 10^{-3} \left(\frac{3~\mathrm{TeV}}{M_{KK}}\right)^2
\frac{f_{-d}/f_{-d1}}{0.22/0.1}, 
\end{array}
\right.
\end{equation}
where, for the numerical estimates, we have assumed that the masses of the
first fermion KK modes are approximately flavor independent $\sim
3$ TeV and we have taken the 
values of the localization parameters from Eq.~(\ref{md:ratios}).
In the usual realization of flavor in models with warped extra
dimensions, the typical value of $C_4^{sd(\mathrm{RS})}$ is about two
orders of magnitude too large to be compatible with experimental
data~\cite{newflavor}. 
Thus, the suppression factor that we obtain seems to be just of
the right order of magnitude to make models with MFP and a low scale
of new physics compatible with flavor constraints.
This suppression assumes that the KK excitations of $q^i$ 
are at least as massive as the KK gluons. This is true for $[++]$ or
$[--]$ boundary conditions but if the corresponding fermions have
twisted ($[+-]$ or $[-+]$) boundary conditions, ultralight modes could
appear. If these ultralight modes 
mix sizably with the RH down quark zero modes, they can induce
too large CP violation in the Kaon system. It should be kept in mind,
however, that although KK quarks much lighter than the gauge boson KK
modes are expected in models of natural EWSB~\cite{Carena:2006bn}, 
they are usually
related to the top and do not necessarily mix significantly with the
RH down sector. 
Furthermore, since the leading contribution in MFP models comes
from a dimension 8 operator, the effect decouples like
$C_4^{sd(MFP)}\sim M_{KK}^{-4}$, as
opposed to models without such protection, for which the decoupling
goes like $C_4^{sd(RS)}\sim M_{KK}^{-2}$. 
In order to better
assess whether this suppression factor is enough or not, a more
detailed numerical analysis is required. The result of such analysis
is discussed in the next section.

\subsection{Effects of Brane Kinetic Terms}

Brane kinetic terms cannot be set to zero at all scales, as they are
generated by quantum corrections~\cite{BKT}. In particular, loops
involving the brane localized Yukawa couplings will generate flavor
non-universal brane kinetic terms for the RH down quarks. The relevant
part of the Lagrangian can be written as
\begin{equation}
\left(\frac{R}{z}\right)^4 \bigg\{
\Big[ \bar{Q}^d_i \left(\cancel{D}+\frac{c}{z}\right) 
Q^d_i \Big]
+\delta(z-R^\prime)R^\prime \Big[  
\bar{Q}^d_i K_{ij} \cancel{D} Q^d_j 
- \frac{v}{\sqrt{2}} 
\big(\bar{q}^{di} \tilde{Y}^d_{ij} [D^j+\tilde{D}^j] +
  \mathrm{h.c.} \big)
\Big]
\bigg\}+ \ldots,
\end{equation} 
where $\cancel{D}$ is the (4d-slashed) covariant derivative (with all
factors of the metric/vielbein stripped out but including the whole
tower of gauge boson KK modes and therefore the full $z$ dependence)
and $K_{ij}$ is the hermitian matrix parametrizing the parallel brane
kinetic terms.~\footnote{We only need to consider parallel brane
  kinetic terms as classical renormalization of the singularities
  due to orthogonal brane kinetic terms bring them to the form of
  parallel brane kinetic terms~\cite{BKT:renormalization}.}
The dots stand for other
fields and $z-dependent$ derivative terms (which are irrelevant for
the discussion here). 
This matrix $K_{ij}$ is in fact \textit{non-calculable} in 5D models, as it
corresponds to a linearly divergent integral. It can be
\textit{estimated}, using NDA to be~\cite{newflavor}
\begin{equation}
K\sim|\tilde{Y}|^2 \frac{\Lambda R^\prime}{16\pi^2},\label{NDA}
\end{equation} 
with $\Lambda$ the
cut-off scale for the Yukawa interaction. 
The hermitian matrix $K$ can
be diagonalized with a unitary rotation, $Q^{di} \to \mathcal{U}_{ij}
Q^{dj}$, such that
\begin{equation}
(\mathcal{U}^\dagger K \mathcal{U})_{ij} = a_i \delta_{ij},
\end{equation}
where $a_i$ are dimensionless numbers of the order of the $K$
entries. The Lagrangian has in the new basis the same form as the one above
(in particular the bulk terms are not affected by this rotation, due
to the MFP prescription) with diagonal BKT's and a new Yukawa
coupling, still order one
\begin{equation}
\tilde{Y}^d \to \mathcal{U}^\dagger \tilde{Y}^d,
\end{equation}
which shows that we can, without loss of generality, start with
diagonal (but in general family non-universal) BKTs.

The effect of diagonal fermion BKTs on the KK expansion was first
studied in the last two references of~\cite{BKT}. 
The result for the zero modes is that they
maintain the same wave function profile as in the case of zero BKT and
only the normalization is modified. The new normalization can be
obtained from the previous one with the replacement
\begin{equation}
f_c \to \frac{f_c}{\sqrt{1+a_i f_c^2}}.
\end{equation}
In particular, the coupling to the first gauge boson KK mode, now
reads, to a very good accuracy,
\begin{equation}
g_{di} \approx g_{s\ast} \left( - \frac{1}{\log R^\prime/R}
+(\gamma(-c_d)+a_i \zeta(-c_d)) \frac{f_{-d}^2}{1+a_i f_{-d}^2}
\right)
= g_d^\mathrm{universal}+g_{s\ast} \zeta(-c_d)a_i f_{-d}^2
+\mathcal{O}(f_{-d}^4),
\end{equation}
where $\zeta(c)$ is an order one function of $c$ and 
in the second equality we have expanded in $f_{-d} \ll 1$.
The large rotation in the $Q^d$ sector will then reintroduce the
dangerous FCNC to the KK
gluons in the RH down sector which are, \textit{a priori}, only
suppressed by the loop factor in $a_i$. Again a full numerical scan,
that we preform in the next section, is
required to asses whether that suppression is enough and whether the
mild hierarchies that are required in the 5D Yukawas to reproduce the
down sector masses, play a role in further suppressing these
contributions.

\section{Numerical Scans\label{section:scan}}

\begin{figure}[t]
\centerline{ 
\includegraphics[width=0.75 \textwidth]{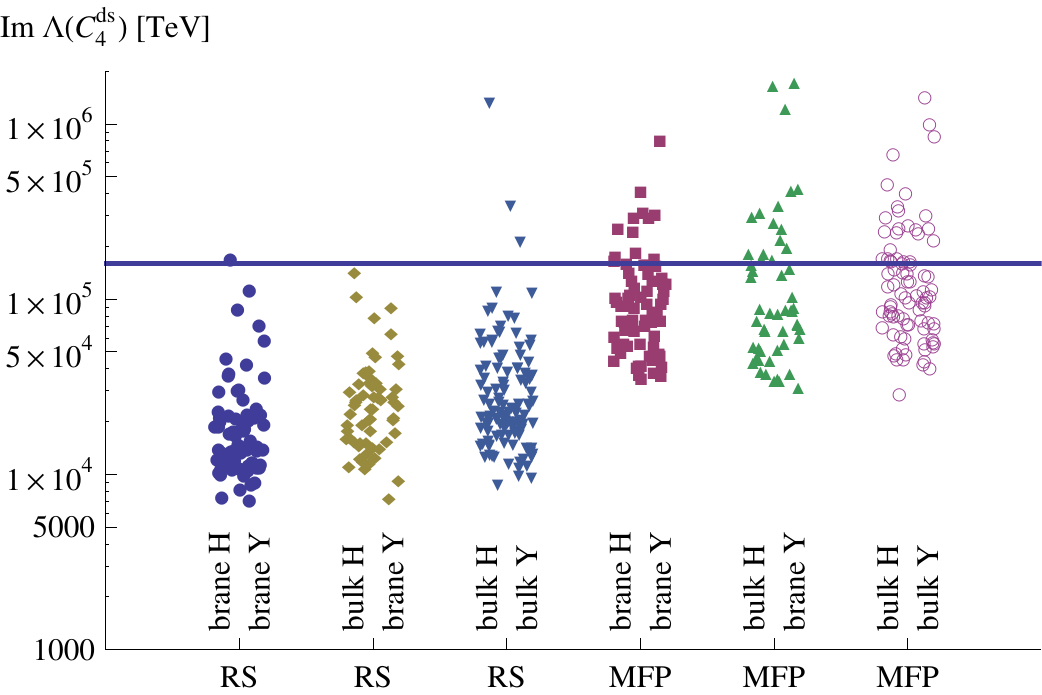}
} 
\caption{Mass suppression of the imaginary part of the 
coefficient $C_4^{sd}$ in TeV for our
  three different models without (left three columns) and with MFP
  (right three columns). In all points we have
  fixed $M_{KK}=3$ TeV and the different sets correspond, from left to
  right, to standard RS with boundary Higgs, RS with bulk Higgs but
  boundary Yukawas, bulk Higgs with bulk up Yukawas and the same three
  models with MFP. The horizontal line corresponds to the experimental 
  lower bound on the suppression scale. A random horizontal
shift in the points has been introduced to facilitate visualization.}
\label{scans}
\end{figure}

The suppression provided by our MFP prescription on the LR
contribution to CP violation in the Kaon system is just about right
(in the absence of large BKT) to
put $\epsilon_K$ around the current experimental limit. In order to be
more quantitative about this bound, we have performed a numerical scan
in which we have chosen different values of $0.4 \leq c_{q3}\leq
0.45$, fixed the values of $c_{q1,q2}$, $c_{ui}$ and $c_{d}$ using
Eqs.~(\ref{fqi}-\ref{md:ratios}), and then generating random complex
$3\times 3$ matrices $\tilde{Y}^{u,d}$, satisfying
$|\tilde{Y}^{u,d}|\leq 3$, in oder to keep perturbativity for the
first few KK modes~\cite{newflavor}. Out of the points generated, we
have selected the ones for which the pattern of quark masses and
mixing angles is similar to the experimentally observed (including the
Jarlskog invariant) and with that subset we have computed the
coefficients of the corresponding gluon mediated $\Delta F=2$
operators. In our numerical studies, we have \textit{not} used the mass
insertion approximation, Eq.~(\ref{mass:insertion}). In order to 
include the effect of fermion mixing to all orders we have
numerically diagonalized the corresponding mass matrices (including
the KK modes of all fermions, not only $q^{(n)i}$) and computed the
couplings to the gluon KK modes in the physical basis using the exact
rotation matrices.
In order to assess separately the two different effects discussed in
the previous section we have not
included fermion 
brane kinetic terms in these scans. They will be analyze in a
more extensive scan described below. Also, for simplicity, we have not
included BKT for the KK gluons.
They can change the bound on the scale
of new physics in either direction by about $50\%$ in the case of
models without MFP~\cite{newflavor}. 
If MFP is at work, the faster decoupling of the
corresponding operators will imply a much smaller variation in the
bound. In order to explore further
different realizations of flavor, we have considered three different
possibilities. The first one is the one we have discussed in the
previous two sections, an IR brane localized Higgs. The other two
possibilities involve a bulk Higgs, with profile in the extra
dimension given by
\begin{equation}
\phi(x,y)=\sqrt{\frac{2}{1-R/R^\prime}} \frac{z}{R^\prime}
\phi^{(0)}(x)+\ldots,
\end{equation}
as motivated by holographic Higgs models~\cite{Contino:2003ve}.
The options now are, to still have only brane Yukawa couplings (despite the
fact that the Higgs is a bulk field) or the more natural option of
having bulk Yukawa couplings
in the up sector, but boundary Yukawas in the down sector (due to the
bulk flavor symmetry). We have realized scans with these three options
both in the case of MFP and in the standard case in which there is no
flavor symmetry. (In this latter case, 
the three different localization parameters in the RH down
sector are computed directly from the mass of the $d,s,b$ quarks.) The
result is displayed in Fig.~\ref{scans}, where a random horizontal
shift in the points has been introduced to facilitate visualization.
In all cases we have fixed $M_{KK}=3$ TeV. The first three sets of
points correspond to RS models with no flavor protection beyond the
RS-GIM mechanism. They are, from left to right, for a boundary Higgs,
bulk Higgs but boundary Yukawas and bulk Higgs with bulk (up) Yukawas,
respectively. The remaining three sets correspond to the same
Higgs and Yukawa configurations but with MFP. It is evident from the plot 
that a value $M_{KK}=3$ TeV in the original RS model
without flavor protection is ruled out by $\epsilon_K$, unless we are
willing to make a fine-tuning of a least $\lesssim 10^{-2}$ 
(corresponding to the ratio of points that survive the bound),
independently of 
where the Higgs and Yukawa couplings live. With MFP, on the
other hand, a sizable region ($\sim 17-30\%$) of parameter space 
is allowed by $\Delta F=2$ observables. Furthermore, due to the extra
suppression with higher values of $M_{KK}$, if we take $M_{KK}=4$ TeV,
the percentage of points above the bound in models with MFP is about
$\sim 40-50\%$, whereas it remains at the per cent level in models
without MFP. 
Finally, although we have only displayed the results for the most
constraining operator $Q_4^{sd}$ in Fig.~\ref{scans}, we have also
checked that all other $\Delta F=2$ operators, including those in the
up sector, are below current
experimental limits~\cite{Bona:2007vi}. 

In order to analyze the impact of BKT for the RH down type quarks, we
have performed some extra very exhaustive scans for the case of a boundary
Higgs. Also, to test the models in a slightly complementary
way to the previous scans, we have fixed the parameters in a different
way. We have generated random complex $3\times 3$ Yukawa matrices
$\tilde{Y}^ {u,d}$, again satisfying $|\tilde{Y}^{u,d}|\leq 3$. We
have then fixed the ratios $f_{qi}/f_{q3}$ and $f_{-ui}/f_{-u3}$, with
$i=1,2$, from Eq.~(\ref{diagonal:masses}) and checked if the resulting 
CKM matrix (including the Jarlskog invariant) agrees with the
experimentally measured
values (better than $30\%$). 
$f_{-u3}$ and $f_{-d}$ 
are then fixed by the top and bottom masses, once we
choose a value for $f_{q3}$. In order to test the dependence on
$c_{q3}$ we have done the analysis with two different values,
$c_{q3}=0.4$ and $c_{q3}=0.45$. In the case that we are not
considering MFP, $f_{-d1}$ and $f_{-d2}$ have been taken randomly
within the ranges suggested by Eqs.~(\ref{diagonal:masses}) and
~(\ref{fqi}). Finally, we have randomly generated 
diagonal BKTs for the RH down quarks
with three different upper bounds, $a_{1,2,3}\leq 3^2 \delta_N$, 
$N=1,2,3$, with
\begin{equation}
\delta_N\equiv\frac{N M_{KK} R^\prime}{16\pi^2}.
\end{equation}
We have done this to estimate the dependence on the
actual size of the BKT. As we mentioned, Eq.~(\ref{NDA}) is nothing
but an estimation, as the actual value of the BKTs cannot be computed
in the five-dimensional theory. Thus, when
we set $N=1$ we do not mean that the cut-off of the model is already
at the first KK mode but rather than the NDA estimation might be a bit
pesimistic and the actual BKT could turn out to be a bit smaller than
expected. 
Similarly, we have also considered the case that the estimation is
accurate and the cut-off is at the third KK mode.

\begin{table}
\begin{tabular*}{0.6\textwidth}{@{\extracolsep{\fill}}lcc}
\hline \hline
& \multicolumn{2}{c}{$\%$ of allowed points} \\
\multicolumn{1}{c}{Model} & $c_{q3}=0.4$ & $c_{q3}=0.45$ \\
\hline
non MFP & 0.4 & 0.4 \\
 MFP & 23.1 & 23.0 \\
 MFP with BKT (N=1) & 11.5 & 14.8 \\
 MFP with BKT (N=2) & 5.9 & 7.8 \\
 MFP with BKT (N=3) & 3.6 & 5.5 
\\ \hline \hline
\end{tabular*}
\caption{\label{results:table}
Ratio of points (in percentage) that are
consistent  with all relevant $\Delta F=2$ constraints for a boundary
Higgs, without MFP, with MFP but no BKT and with MFP and BKT with
three different values for the maximun allowed size of the BKT. In
these scans we have only included points that satisfy
$\mathrm{Max}(|\tilde{Y}^{u,d}|) /\mathrm{Min}(|\tilde{Y}^{u,d}|) \leq
10$ (see text for details).}
\end{table}

We have generated several thousand points along the previous lines and
have analyzed the constraints from all relevant $\Delta F=2$
observables~\cite{Bona:2007vi}. The results are summarized in
Table~\ref{results:table}, where we report on the percentage of points
that pass \textit{all} $\Delta F=2$ constraints for each model and
size of BKT. These results show a very good consistency with our
previous scans in the case that BKTs are neglected (both with and without
MFP). In the case of MFP with BKTs, the ratio of points that
pass all $\Delta F=2$ constraints is reduced as expected. 
However, we find that the level of
reduction is very sensitive to the size of the BKTs and also somewhat
sensitive to the exact localization of the LH top/bottom quarks, 
$c_{q3}$. In particular, for
$c_{q3}=0.45$ and $N=1$ (BKT a bit smaller than the NDA estimation),
about $15\%$ of the points are allowed by flavor constraints, getting
reduced to $4\%$ in the case of $c_{q3}=0.4$ and $N=3$, smaller but
still sizably better than the case without MFP.
These numbers have been obtained including only those points that do not
involve a hierarchy between the different entries of the Yukawa
matrices larger than a factor of 10,
$\mathrm{Max}(|\tilde{Y}^{u,d}|) /\mathrm{Min}(|\tilde{Y}^{u,d}|) \leq
10$, but we have checked that the results are not very sensitive to
that factor. For instance they change by a few per cent up to $\sim
50\%$ for the case of the models with BKT and $N=3$ if we use only
the points with a hierarchy in the Yukawas smaller than 5 (although
the statistics is much limited in that case) and barely change if we
allow an arbitrary hierarchy.

\section{Discussion\label{discussion}}

Recent analyses have shown that flavor observables strongly constrain
the parameter space of models with warped extra dimensions and a
considerable effort is being put in finding successful flavor
symmetries that protect these models from large flavor violations. 
We have argued that the status of flavor in models with
warped extra dimensions is actually not that bad and current
experimental limits do not yet require desperate measures. In fact,
the built-in flavor protection mechanism present
in models with warped extra dimensions, the RS-GIM mechanism, which
ensures most of the flavor violating observables to remain close to but
below current experimental bounds works extremely well. It is only
one observable (CP violation in Kaon oscillations) 
and only due to a very particular,
chirality enhanced, contribution that significantly constrains the
parameter space in these models. Thus, we have suggested a very simple
solution that suppresses the only dangerous set of operators, leaving
intact most of the rest of the flavor violating and diagonal structure
of the models. 
In this work, we have shown that the proposal can be
easily realized in the quark sector, by imposing a flavor symmetry
that enforces a common localization parameter for all the RH quark zero modes
in the down sector. We have shown that the contribution to $\Delta
F=2$ operators with RH currents (including the chirality enhanced LR
contributions) is vanishing at leading (tree level dimension 6 operator)
order. We have then computed the most relevant 
sub-leading (dimension 8 and some loop suppressed) 
contributions to $\Delta F=2$ processes.
A numerical scan shows that, including the contributions from
dimension 8 operators,
a significant region of parameter space, with a very low KK
scale ($M_{KK}=3-4$ TeV) is compatible with flavor and electroweak
precision data. 
The presence of (loop suppressed) brane kinetic terms, reintroduce
FCNC in the RH down sector, thus representing the leading source of
flavor violation in the down sector, if they are large. We have
performed dedicated scans to analyze the effect of BKTs with the
result that they certainly reduce the allowed region of parameter
space if they are large. However, we have seen that, if the BKTs are
just a bit smaller than the NDA \textit{estimate}, their effect is not
dramatic and we 
still have sizable regions of parameter space allowed by flavor and
electroweak precision tests. 
In this allowed regions, the hierarchies in the quark
masses and mixing angles are still naturally explained through wave
function localization. In fact, MFP provides a rationale for the
smaller hierarchy in the down sector. This shows that the MFP
idea can lead to viable models with warped extra dimensions and new
physics accessible at the LHC.

In order to fully study the viability of MFP as
a new paradigm in models with warped extra dimensions, some further
analyses, some of which are currently under way, are required. First,
we have considered the more constraining $\Delta F=2$
observables, including the contribution of the first few KK
modes. This gives the main contribution, but a more refined
analysis, including the effect of the whole tower of gauge boson and
fermion KK modes, which can be done analytically in some
cases~\cite{newflavor,neubert}, would be desirable.
Also, a detailed analysis of $\Delta F=1$ and electric dipole moment
(EDM) contributions can decide which observables are the most likely
to give experimental signatures in minimal flavor protection
scenarios. $\Delta F=1$ processes were shown in~\cite{Agashe:2004cp}, 
in the absence of flavor
protection, to give contributions close to but below current
experimental limits, in particular in semileptonic decays of $B$
mesons. Also, the authors of~\cite{Agashe:2004cp} showed that the neutron
EDM generates a new strong CP problem, requiring a cancellation of
about $5\%$. It would be interesting to investigate
the situation in the presence of MFP, which, at least for two
generations, seems to provide enough suppression of the neutron EDM.
Also, we have not explicitly studied models of gauge-Higgs
unification. These were shown in~\cite{newflavor} to be slightly more
constraining than the standard RS model with a fundamental Higgs, due
to a kinetic mixing induced by the localized couplings needed to
generate non-trivial quark masses and mixings. The resulting
bound on $M_{KK}$ turns out to be $\sim 50\%$ stronger than in the
case of a fundamental Higgs. In our case, due to the extra power of
$M_{KK}^{-2}$ on the correction, a similar result as the one displayed
in Fig.~\ref{scans}, can be expected for gauge-Higgs unification
models with $M_{KK}\sim 4$ TeV, provided no ultralight modes mix
sizably with $d_R^{(0)i}$. A detailed analysis, would be however
required to fully assess the effectiveness of the MFP paradigm in
models of gauge Higgs unification.  

Finally, it would be interesting to study how the MFP paradigm can be
extended to the leptonic sector. The smaller (about three orders of
magnitude) hierarchy of charged lepton masses and the large neutrino
mixing angles seem to point to symmetries enforcing
universality of some of the 
localization parameters in the lepton sector. It would
be interesting to analyze the interplay of lepton masses and mixing
angles and lepton flavor violating processes~\cite{Agashe:2006iy} 
to try and find the
minimal flavor protection scheme in the leptonic sector.

In summary, we have argued that the current experimental situation of
models with warped extra dimensions does not require extreme measures
yet. Minimal flavor protection on top of the RS-GIM mechanism,
together with custodial symmetry and a symmetry protection of the $Z
\bar{b}_L b_L$ coupling seem enough to provide fully realistic models
of warped extra dimensions accessible at the LHC. The beauty of 
minimal flavor protection, beyond its simplicity, is that it still 
predicts flavor violating processes close to current experimental
limits and can therefore be also tested with precision flavor
experiments. If these experiments end up ruling out the minimal flavor
protection paradigm, less minimal flavor constructions along the lines
of~\cite{Cacciapaglia:2007fw} would be required.

\vspace{12pt}
\noindent\textbf{Note Added:} While this work was being completed,
Refs.~\cite{note} became public. Two different approaches
to flavor symmetries in the leptonic sector are discussed in these
works.  In both works universalities of localization parameters play
an important role to generate large neutrino mixing angles and prevent
excessive flavor violation along the lines mentioned above. 

\section*{Acknowledgments} We would like to thank 
K. Agashe, C. Csaki, A. Delgado, A. Falkowski, U. Haisch, 
M. Neubert, E. Pont\'on, T. Tait and A. Weiler for useful 
discussions and comments. Financial support from the Swiss National
Science Foundation under contract
200021-117873 is acknowledged. We would also like to thank B. Menor and
C. Menor for creating a wonderful working environment during the final
stages of this work.

\end{document}